\documentclass[letterpaper, 10 pt, conference]{ieeeconf}  

\IEEEoverridecommandlockouts                              

\usepackage{graphics} 
\usepackage{epsfig} 
\usepackage{amsmath} 
\usepackage{amssymb}  
\usepackage{bm}
\usepackage{booktabs}
\usepackage{multirow}
\usepackage{color}

\title{\LARGE \bf
Fast MRI Reconstruction: How Powerful Transformers Are?
}

\author{Jiahao Huang$^{1,2,\dagger}$, Yinzhe Wu$^{1,3}$, Huanjun Wu$^{1,3}$, Guang Yang$^{1,2,\dagger}$
\thanks{$^{1}$National Heart and Lung Institute, Imperial College London, London, SW7 2AZ, United Kingdom}%
\thanks{$^{2}$Cardiovascular Research Centre, Royal Brompton Hospital, London, SW3 6NP, United Kingdom}%
\thanks{$^{3}$Department of Bioengineering, Imperial College London, London, SW7 2AZ, United
Kingdom}%
\thanks{$^{\dagger}$Send correspondence to \{j.huang21,g.yang\}@imperial.ac.uk}%
\thanks{This study was supported in part by the BHF (TG/18/5/34111, PG/16/78/32402), the ERC IMI (101005122), the H2020 (952172), the MRC (MC/PC/21013), and the UKRI Future Leaders Fellowship (MR/V023799/1).}%
}

\begin{document}

\maketitle
\thispagestyle{empty}
\pagestyle{empty}

\begin{abstract}

Magnetic resonance imaging (MRI) is a widely used non-radiative and non-invasive method for clinical interrogation of organ structures and metabolism, with an inherently long scanning time. Methods by \textit{k}-space undersampling and deep learning based reconstruction have been popularised to accelerate the scanning process.
This work focuses on investigating how powerful transformers are for fast MRI by exploiting and comparing different novel network architectures. 
In particular, a generative adversarial network (GAN) based Swin transformer (ST-GAN) was introduced for the fast MRI reconstruction. To further preserve the edge and texture information, edge enhanced GAN based Swin transformer (EES-GAN) and texture enhanced GAN based Swin transformer (TES-GAN) were also developed, where a dual-discriminator GAN structure was applied.
We compared our proposed GAN based transformers, standalone Swin transformer and other convolutional neural networks based GAN model in terms of the evaluation metrics PSNR, SSIM and FID.
We showed that transformers work well for the MRI reconstruction from different undersampling conditions. The utilisation of GAN's adversarial structure improves the quality of images reconstructed when undersampled for 30\% or higher. The code is publicly available at https://github.com/ayanglab/SwinGANMR.

\end{abstract}

\section{INTRODUCTION}
Magnetic resonance imaging (MRI) is a widely used non-radiative and non-invasive method for interrogation of organ structures and metabolism~\cite{chen2021ai}. However, a fully sampled MRI with high spatial resolution may require a long time to acquire~\cite{Zbontar2018}. Despite endeavours in parallel imaging and compressive sensing, traditional fast MRI methods suffered from a limited acceleration factor and a prolonged iterative procedure~\cite{yang2021generative}.  

Recently, convolutional neural network (CNN) based models~\cite{Wang2016,schlemper2018stochastic,lv2021transfer} were developed for the post-acquisition reconstruction of undersampled MRI that leveraged their hierarchical structures to establish the latent sparse correlations in both \textit{k}-space and image space between the undersampled and fully sampled MR images.

\begin{figure}[thpb]
  \centering
  \includegraphics[scale=0.62]{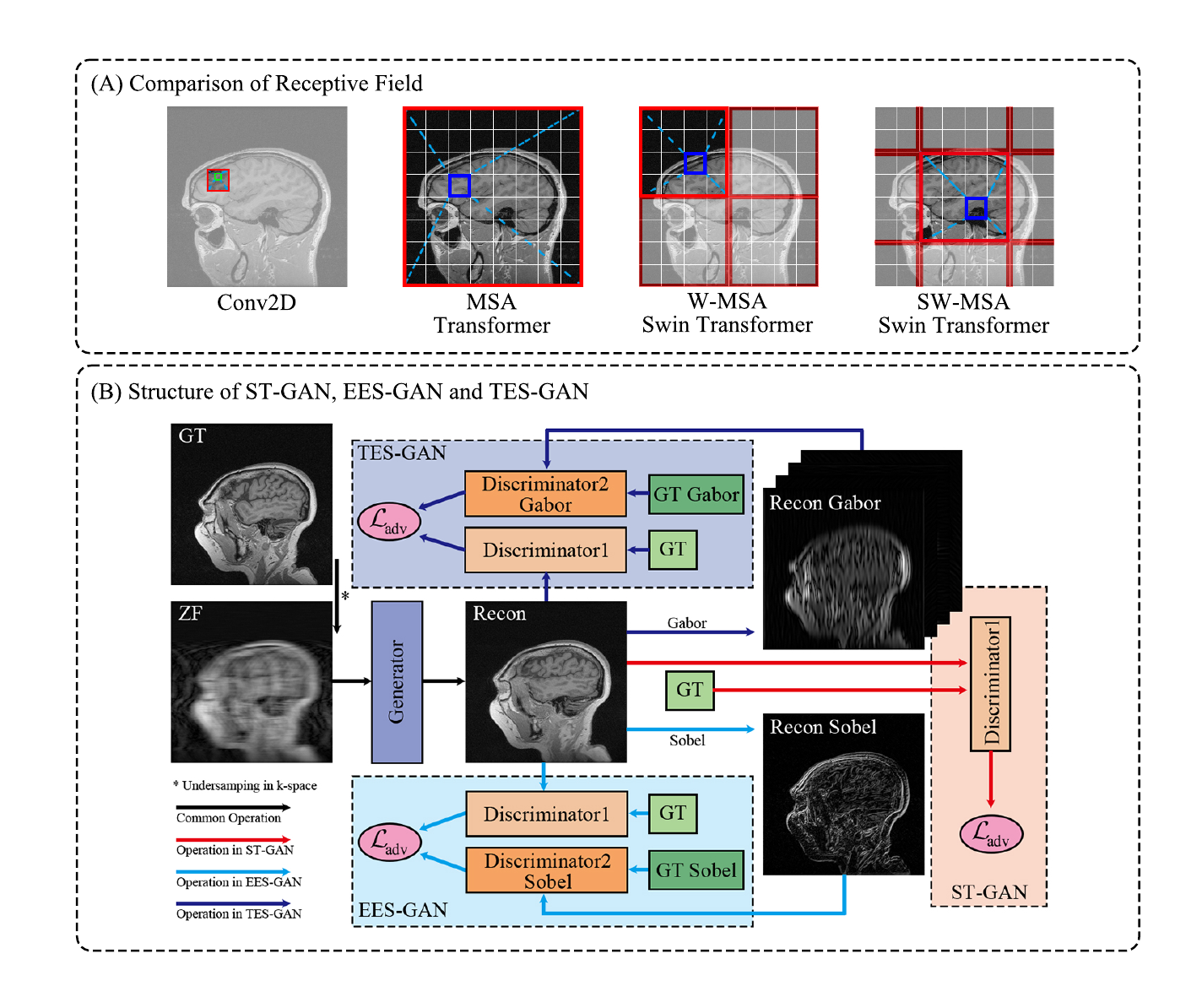}
  \caption{
  (A) The schematic diagrams of receptive fields for 2D convolution (Conv2D), multi-head self-attention (MSA) in vanilla transformer, windows based multi-head self-attention (W-MSA) and shifted windows based multi-head self-attention (SW-MSA) in \textbf{S}hifted \textbf{win}dows (Swin) transformer.
  Red box: Receptive fields; Green box: Pixels; Blue box: Patches.
  (B) The architecture of the proposed single-discriminator GAN based Swin transformer (ST-GAN), dual-discriminator edge enhanced GAN based Swin transformer (EES-GAN) and dual-discriminator texture enhanced GAN based Swin transformer (TES-GAN).
  }
  \label{fig:overview}
\end{figure}

More recently, models with expanded receptive fields~\cite{Parmar2018}, namely transformers, started gaining attention, for their unique advantage through its sequence-to-sequence model design~\cite{Sutskever2014} and adaptive self-attention setting~\cite{Matsoukas2021}. 
Instead of spanning receptive fields across the whole image, a variant of transformer with moving receptive field windows of reduced sizes was proposed, namely \textbf{S}hifted \textbf{win}dows (Swin) transformer~\cite{Liu2021}, as Fig.~\ref{fig:overview} (A) shown. Such design improved the adaptability of the transformer based models, while greatly reducing the computational complexity.

In this work, we focus on investigating how powerful transformers are for fast MRI by exploiting and comparing different novel architectures. In particular, the Swin transformer based GAN (ST-GAN) model (Fig.~\ref{fig:overview} (B)) is proposed for the fast MRI reconstruction, with a Swin transformer based generator and a discriminator for holistic MR image reconstruction. 
Besides, inspired by the dual-discriminator GAN structure for edge and texture preservation, edge enhanced GAN based Swin transformer (EES-GAN) and texture enhanced GAN based Swin transformer (TES-GAN) are also developed, to further exploit the combination of the transformers and GAN structure for MRI reconstruction. We compare these novel transformer based GANs and the standalone Swin transformer with other CNN based GANs and zero-filled baselines.

\section{METHOD}

\subsection{Formulation}

MRI reconstruction aims to recover latent images $x$ from the undersampled \textit{k}-space measurements $y$. Traditionally, the reconstruction problem can be converted into a optimisation problem as follows
\begin{align}\label{formula:mri_recon}
\mathop{\text{min}}\limits_{x} \frac{1}{2} 
\mid\mid y - \mathcal{M}\mathcal{F} x\mid\mid^2_2
+ \lambda R(x),
\end{align}
\noindent where $R(x)$ is the regularisation term balanced by $\lambda$ and $\mid\mid \cdot \mid\mid_2$ denotes the $l_2$ norm. $\mathcal{M}$ and $\mathcal{F}$ denote the undersampling trajectory and Fourier transform.

With its superior ability of feature extraction, CNN has been applied to MRI reconstruction to reduce the long reconstruction time of traditional methods~\cite{Schlemper2018}, which can be formulated by
\begin{align}\label{formula:cnn_mri}
\min_{x} \frac{1}{2} 
\mid\mid y - \mathcal{M}\mathcal{F} x\mid\mid^2_2
+ \lambda \mid\mid x - f_{\mathrm{CNN}}(x_u\mid\theta) \mid\mid^2_2,
\end{align}
\noindent where $f_{\mathrm{CNN}}$ is the CNN trained to map undersampled MR images $x_u$ to reconstructed MR images $\hat x_u$.

\subsection{Network Architecture}

\subsubsection{Swin Transformer Based Generator}
\begin{figure}[thpb]
  \centering
  \includegraphics[scale=0.6]{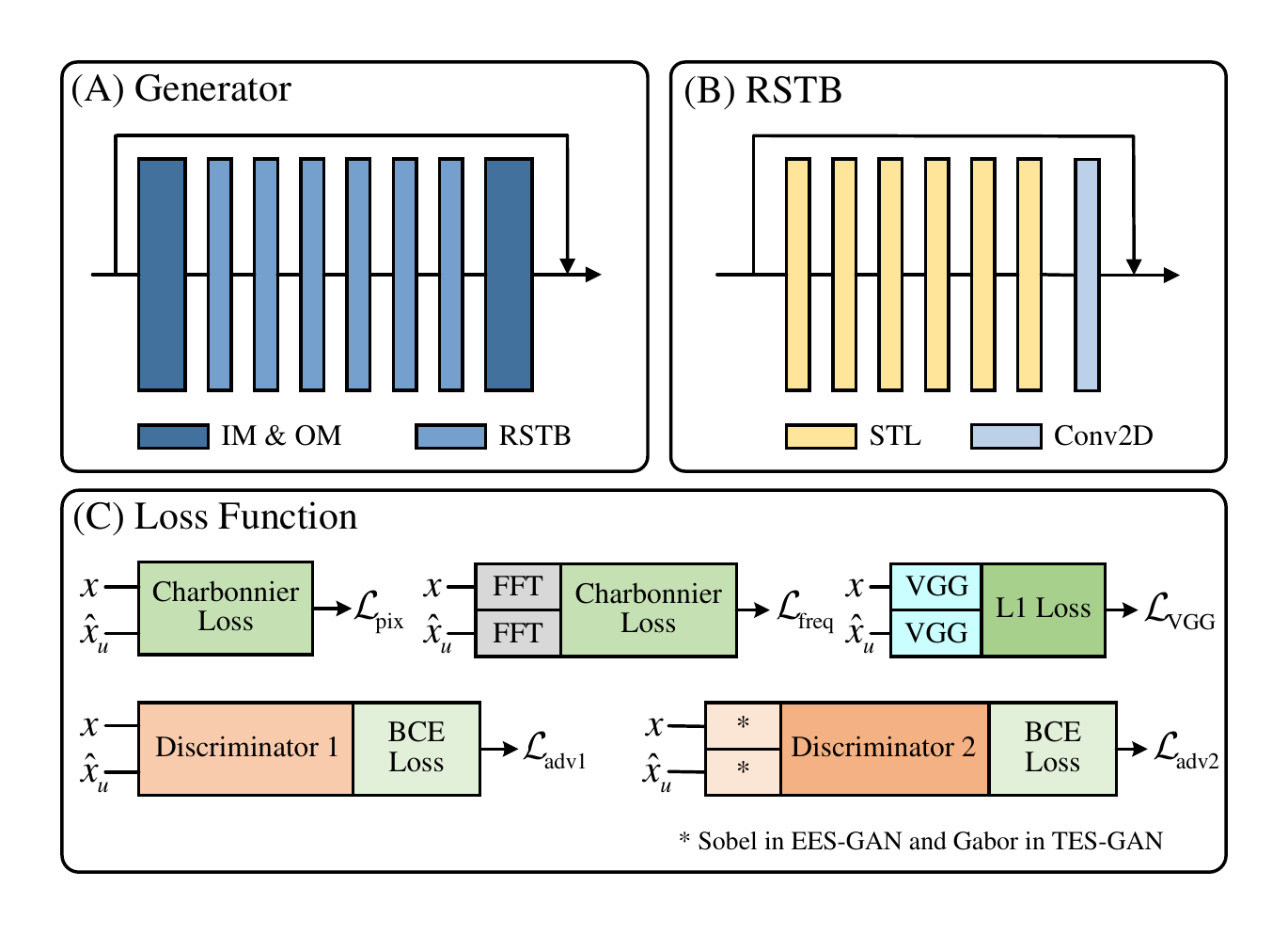}
  \caption{
  (A) Structure of the Swin transformer based generator. IM: input module; OM: output module. 
  (B) Structure of the residual Swin transformer blocks (RSTBs). STL: Swin transformer layer; Conv2D: 2D convolutional layer. 
  (C) Loss functions applied.
  $\mathcal{L}_{\mathrm{pix}}$: Pixel-wise Charbonnier loss; 
  $\mathcal{L}_{\mathrm{freq}}$: Frequency Charbonnier loss; 
  $\mathcal{L}_{\mathrm{VGG}}$: Perceptual VGG L1 loss; 
  $\mathcal{L}_{\mathrm{adv1}}$: Adversarial loss from discriminator for holistic image reconstruction (in ST-GAN, EES-GAN and TES-GAN);
  $\mathcal{L}_{\mathrm{adv2}}$: Adversarial loss from discriminator for edge and texture information preservation (in EES-GAN and TES-GAN).
  }
  \label{fig:structure}
\end{figure}

As Fig.~\ref{fig:structure} (A) shows, a Swin transformer based generator~\cite{Liang2021, Huang2022}, which consists of an input module (IM), a cascaded of residual Swin transformer blocks (RSTBs) and an output module (OM), is applied in our proposed ST-GAN, EES-GAN and TES-GAN. 
A residual connection is applied between the input and output to stable the training process as followed:
\begin{align}\label{formula:refinement}
\hat x_u = G_{\theta_G}(x_u) + x_u.
\end{align}
The IM is a Conv2D at the beginning of the network for the shallow feature extraction, which maps the images space $\mathbb{R}^{h \times w \times 1}$ to high dimension feature space $\mathbb{R}^{H \times W \times C}$. The channel is enlarged for the follow-up transformer module ($h=H, w=W$).
The OM is a Conv2D placed at the end, mapping from the high dimension feature space $\mathbb{R}^{H \times W \times C}$ to the output image space $\mathbb{R}^{h \times w \times 1}$.

As Fig.~\ref{fig:structure} (A) shows, the RSTB is composed of a series of Swin transformer layers (STLs) and a Conv2D with a residual connection. Patch embedding and patch unembedding operations are placed before the first STL and after the Conv2D to convert the feature map between $\mathbb{R}^{H \times W \times C}$ and $\mathbb{R}^{HW \times C}$.

\subsubsection{Edge and Texture Enhanced GAN Structure}

As Fig.~\ref{fig:overview} (B) shows, in our proposed ST-GAN, standard two-player GAN structure, i.e, one generator and one discriminator is applied. The only discriminator $D_{\theta_{D_1}}$ is a U-Net based discriminator~\cite{Schonfeld2020} for the holistic images reconstruction, which aims to distinguish reconstructed MR images $\hat x_u$ from ground truth MR images $x$.

For the proposed EES-GAN and TES-GAN, dual-discriminator GAN structures were utilised to train the generator. Similar to ST-GAN, the discriminator for holistic reconstruction $D_{\theta_{D_1}}$ also takes the reconstructed MR images $\hat x_u$ and the ground truth MR images $x$ as input.
In the proposed EES-GAN, an additional U-Net based discriminator for edge information preservation $D_{\theta_{D_2}}$ takes the edge information of both $\mathcal{S}(x)$ and $\mathcal{S}(\hat x_u)$ extracted by the Sobel operator $\mathcal{S}(\cdot)$ as the input. 
In the proposed TES-GAN, an additional U-Net based discriminator for texture information preservation is applied, whose inputs are the texture information of both $\mathcal{G}(x)$ and $\mathcal{G}(\hat x_u)$ extracted by the Gabor operator $\mathcal{G}(\cdot)$.

\subsection{Loss Function}

As Fig.~\ref{fig:structure} (C) shown, the loss function consists of a pixel-wise loss $\mathcal{L}_{\mathrm{pix}}$, a frequency loss $\mathcal{L}_{\mathrm{freq}}$, a perceptual VGG loss $\mathcal{L}_{\mathrm{VGG}}$ and an total adversarial loss $\mathcal{L}_{\mathrm{adv}}$.

The pixel-wise loss and the frequency loss are defined as
\begin{align}\label{}
\mathop{\text{min}}\limits_{\theta_G} 
\mathcal{L}_{\mathrm{pix}}({\theta_G} ) =
\sqrt{\mid\mid x - \hat x_u \mid\mid^2_2 + \epsilon^2},
\end{align}
\begin{align}\label{}
\mathop{\text{min}}\limits_{\theta_G} 
\mathcal{L}_{\mathrm{freq}}({\theta_G} ) =
\sqrt{\mid\mid y - \mathcal{F} \hat x_u \mid\mid^2_2 + \epsilon^2},
\end{align}
\noindent where Charbonnier loss~\cite{Lai2019} is utilised for its superior robustness for outliers, and $\epsilon$ is empirically set to $10^{-9}$. 

The perceptual VGG loss measures the high-dimension mapping between two images by a pre-trained VGG network, which is defined as 
\begin{align}\label{}
\mathop{\text{min}}\limits_{\theta_G} 
\mathcal{L}_{\mathrm{VGG}}({\theta_G} ) =
\mid\mid f_{\mathrm{VGG}}(x) - f_{\mathrm{VGG}}(\hat x_u) \mid\mid_1,
\end{align}
\noindent where $f_{\mathrm{VGG}}(\cdot)$ denotes the pre-trained VGG network and $\mid\mid \cdot \mid\mid_1$ is the $l_1$ norm. 

For ST-GAN, the total adversarial loss is defined as
\begin{align}\label{}
&\mathop{\text{min}}\limits_{\theta_G} 
\mathop{\text{max}}\limits_{\theta_{D_1}}
\mathcal{L}(\theta_G, \theta_{D_1})
\\&=\mathbb{E}_{x \sim p_{\mathrm{t}}(x)}  \nonumber
[\mathop{\text{log}} D_{\theta_{D_1}}(x)]
- \mathbb{E}_{x_u \sim p_{\mathrm{u}}(x_u)}
[\mathop{\text{log}} D_{\theta_{D_1}}(\hat x_u)],
\end{align}
\noindent where $G_{\theta_{G}}$ and $D_{\theta_{D_1}}$ denote the generator and the discriminator for holistic images reconstruction.

For EES-GAN and TES-GAN, the total adversarial loss is defined as
\begin{align}\label{}
&\mathop{\text{min}}\limits_{\theta_G} 
\mathop{\text{max}}\limits_{\theta_{D_1}}
\mathop{\text{max}}\limits_{\theta_{D_2}}
\mathcal{L}_{\mathrm{adv}}(\theta_G, \theta_{D_1}, \theta_{D_2})\nonumber
\\&=\mu \mathcal{L}_{\mathrm{adv1}}(\theta_G, \theta_{D_1}) + \nu \mathcal{L}_{\mathrm{adv2}}(\theta_G, \theta_{D_2})  \nonumber
\\&=\mu\{\mathbb{E}_{{x} \sim p_{\mathrm{t}}({x})}
[\mathop{\text{log}} D_{\theta_{D_1}}({x})] \nonumber
\\&-\mathbb{E}_{{x_u} \sim p_{\mathrm{u}}({x_u})}
[\mathop{\text{log}} D_{\theta_{D_1}}(\hat x_u)]\} \nonumber
\\&+\nu\{\mathbb{E}_{{x} \sim p_{\mathrm{t}}({x})}
[\mathop{\text{log}} D_{\theta_{D_2}}(\mathcal{A}({x}))] \nonumber
\\&-\mathbb{E}_{{x_u} \sim p_{\mathrm{u}}({x_u})}
[\mathop{\text{log}} D_{\theta_{D_2}}(\mathcal{A}(\hat x_u))]\}, 
\end{align}
\noindent where $G_{\theta_{G}}$, $D_{\theta_{D_1}}$ and $D_{\theta_{D_2}}$ denote the generator, the discriminator for holistic images reconstruction and the discriminator for edge and texture information preservation, respectively. $\mathcal{A}(\cdot)$ denotes Sobel operator $\mathcal{S}(\cdot)$ in EES-GAN and Gabor $\mathcal{G}(\cdot)$ operator in TES-GAN. $\mu$ and $\nu$ are the coefficient that balanced two terms.

The total loss can be formulated by 
\begin{align}\label{}
\mathcal{L}_{\mathrm{TOTAL}}
= \alpha \mathcal{L}_{\mathrm{pix}}
+ \beta \mathcal{L}_{\mathrm{freq}}
+ \gamma \mathcal{L}_{\mathrm{VGG}}
+ \delta \mathcal{L}_{\mathrm{adv}},
\end{align}
\noindent where $\alpha$, $\beta$, $\gamma$ and $\delta$ are the coefficients that balance each term.

\section{EXPERIMENTS AND RESULTS}

\subsection{Dataset}

Our proposed methods were trained and tested on the Calgary Campinas dataset~\cite{Souza2018}, which contains 47 cases of 12-channel T1-weight brain MR images.
We randomly chose 40 cases for training, 7 cases for validation and 20 cases for testing, according to the ratio of 6:1:3 approximately. For each case 100 2D sagittal-view slices were chosen.

\subsection{Implement Details}

The proposed ST-GAN, EES-GAN and TES-GAN were trained on two NVIDIA RTX 3090 GPU with 24GB GPU RAM and tested on an NVIDIA RTX 3090 GPU or an Intel Core i9-10980XE CPU.
We applied 6 RSTBs and 6 STLs in each RSTB in the generator, and the patch and channel number were set to 96 and 180 respectively.
The parameters in the total loss function $\alpha$, $\beta$, $\gamma$ and $\delta$ were set to 15, 0.1, 0.0025 and 0.1, respectively. For the EES-GAN and TES-GAN, $\mu$ and $\nu$ were set to 0.05 and 0.05.

\subsection{Evaluation Methods}

In the experiment section, Peak Signal-to-Noise Ratio (PSNR), Structural Similarity Index Measure (SSIM), and Fr\'echet Inception Distance (FID)~\cite{Heusel2017} were applied for the evaluation of different methods.
PSNR is a shallow pixel-scale evaluation metric that presents the ratio between maximum signal power and noise power of two images. 
SSIM is a shallow perceptual based evaluation metric measuring the structural similarity between two images.
FID calculates the Fr\'echet distance between image sets by using a pre-trained Inception V3 network, measuring the similarity between two image sets.

PSNR and SSIM are not sufficient for measuring the visual perceptional quality of images, since the visual perceptional experience is subject to more latent relationships in a higher dimension~\cite{Zhang2018}, where FID is correlated well and more appropriate here.

The number of parameters (\#PARAMs) and Multiply-Accumulate Operations (MACs) were applied to measure the model size and the computational cost. MACs were calculated using a $1\times1\times256\times256$ array as input (Batch $\times$ Channel $\times$ Height $\times$ Width).

\subsection{Comparisons with Other Methods}

In this experiment, our proposed ST-GAN, EES-GAN and TES-GAN were compared with other MRI reconstruction methods, including CNN based GANs method, i.e., DAGAN~\cite{Yang2018} and PIDDGAN~\cite{Huang2021}, and Swin transformer based method, i.e., SwinMR~\cite{Huang2022} using Gaussian 1D 10\% (G1D10\%) and 30\% (G1D30\%) \textit{k}-space undersampling mask. 

\begin{figure}[thpb]
  \centering
  \includegraphics[scale=0.20]{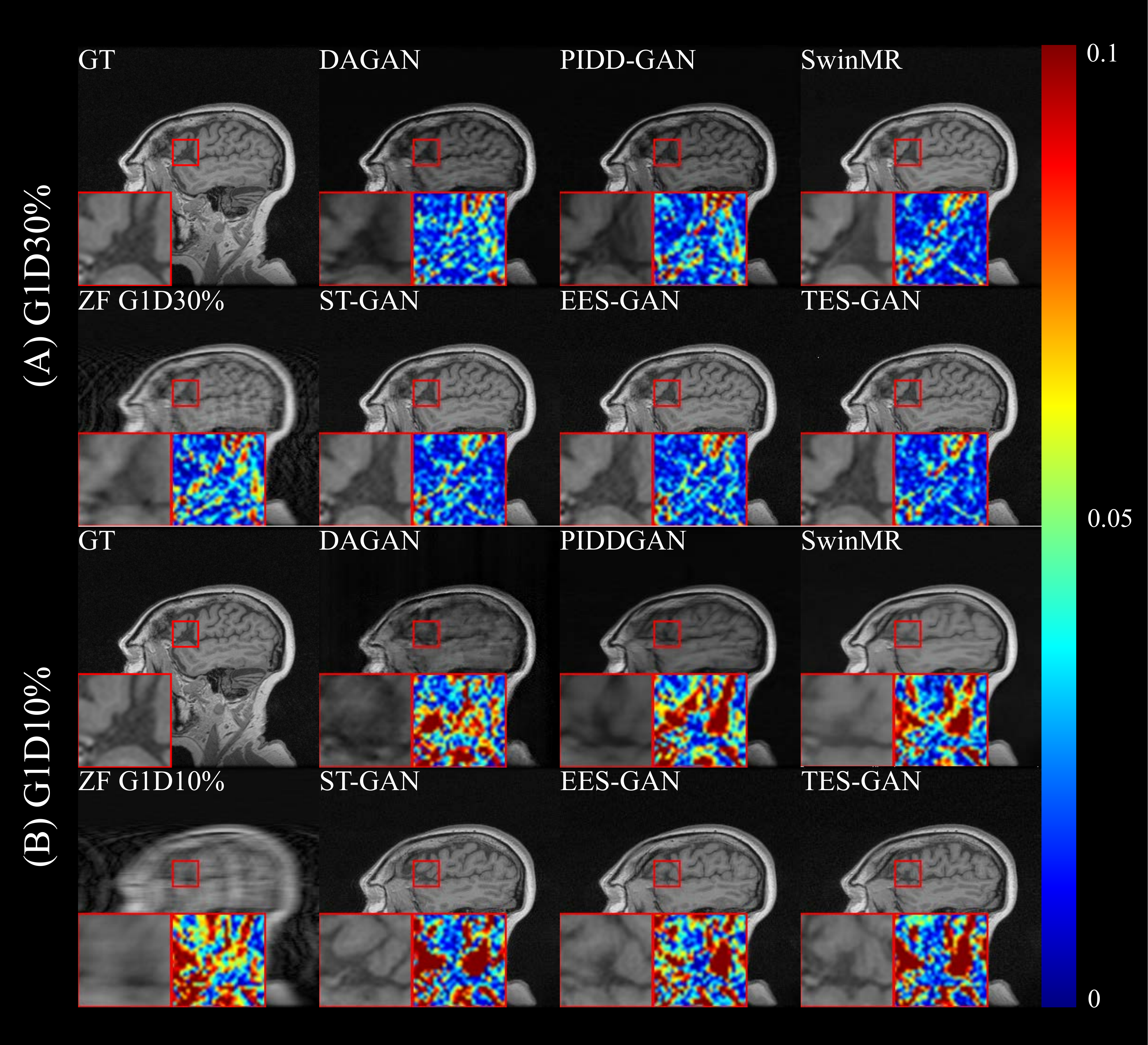}
  \caption{
  Comparison results of different models (DAGAN, PIDD-GAN, SwinMR, ST-GAN, EES-GAN and TES-GAN) against the fully sampled ground truth MR images (GT), with reference to the undersampled zero-filled MR images (ZF), under different sampling rates of (A) 10\% and (B) 30\% by Gaussian 1D (G1D) undersampling masks.
  }
  \label{fig:comparison_sample}
\end{figure}

\begin{figure}[thpb]
  \centering
  \includegraphics[scale=0.55]{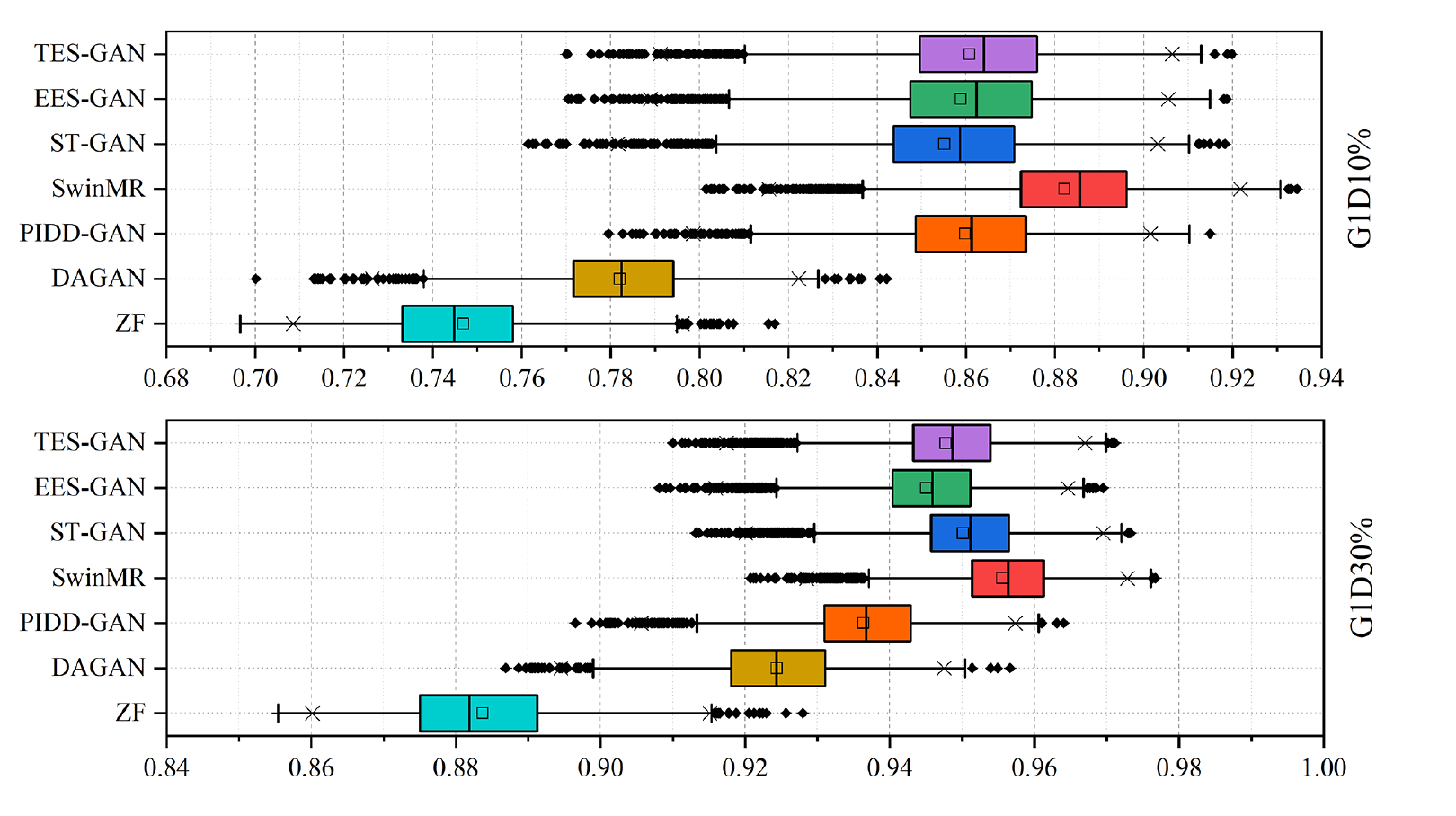}
  \caption{
    Boxplots illustrating the distributions of Structural Similarity Index Measure (SSIM) of the results from different models against the fully sampled ground truth (GT), with reference to the pre-reconstruction undersampled zero-filled MR images (ZF) using Gaussian 1D 10\% and 30\% masks.
    Paired t-test has been performed between methods, and the difference in distribution between any two groups is significant ($p < 0.05$). 
    (Box range: interquartile range; $\times$:1\% and 99\% confidence interval; $-$: maximum and minimum; $\square$: mean; $\shortmid$: median; $\blacklozenge$: outlier.)
    }
  \label{fig:SSIM}
\end{figure}

Fig.~\ref{fig:comparison_sample} displays the samples of the ground truth (GT), undersampled zero-filled images (ZF) and the reconstructed MR images with G1D10\% and G1D30\% masks. 
Fig.~\ref{fig:SSIM} and TABLE~\ref{tab:FID} show the quantitative results of reconstruction by different methods. More results are placed in Supplementary.

\begin{table}[htbp]
  \centering
  \caption{
  \'Frechet Inception Distance (FID), the times for training and inference, number of parameters (\#PARAMs) and Multiply-Accumulate Operations (MACs) using different models, with reference to the undersampled zero-filled images (ZF) using Gaussian 1D 10\% and 30\% masks.\\ $^*$: \#PARAMs for only the generator/ both the generator and discriminators. $\star$: training with different batch sizes and a fair comparison can't be presented.
  }
  \scalebox{0.8}{
    \begin{tabular}{ccccccc}
    \toprule
    \multirow{2}[4]{*}{Method} & \multicolumn{2}{c}{FID} & \multicolumn{2}{c}{Time (s)} & \#PARAMs & MACs \\
\cmidrule{2-5}          & G1D10\% & G1D30\% & Train & Inference & (M)   & (G) \\
    \midrule
    ZF    & 325.99 & 156.38 & -     & -     & -     & - \\
    DAGAN & 169.82 & 56.04 & $\star$     & \textbf{0.003} & 98.59/127.18$^*$ & 33.97 \\
    PIDD-GAN & 90.75 & 17.55 & $\star$     & 0.006 & 32.31/89.50$^*$ & 56.44 \\
    SwinMR & 60.60 & 21.03 & \textbf{59.269} & 0.041 & 11.40 & 800.73 \\
    \midrule
    ST-GAN & 18.86 & 8.50  & 85.921 & 0.041 & 11.40/87.88$^*$ & 800.73 \\
    EES-GAN & 20.50 & 8.48  & 114.028 & 0.041 & 11.40/164.36$^*$ & 800.73 \\
    TES-GAN & \textbf{17.98} & \textbf{7.60} & 114.703 & 0.041 & 11.40/164.37$^*$ & 800.73 \\
    \bottomrule
    \end{tabular}%
    }
  \label{tab:FID}%
\end{table}%

The adversarially trained transformers, i.e., ST-GAN, EES-GAN and TES-GAN, can produce richer edge and texture details in images upon reconstruction, particularly for G1D10\% zero-filled images. However, these enhanced texture details could be wrongly represented (Fig.~\ref{fig:comparison_sample} (B)). Fig.~\ref{fig:comparison_sample} also shows that the extra discriminator in dual-discriminator GANs (EES-GAN and TES-GAN) improved the perceptual quality of the reconstruction but not obviously.

SwinMR achieved the results with the highest SSIM (SwinMR/ST-GAN: 0.882/0.855 by G1D10\%) and PSNR (27.68/26.82), whereas the SSIMs and PSNRs of the results reconstructed by all transformer based GANs (ST-GAN, EES-GAN, TES-GAN) fell slightly behind (Fig.~\ref{fig:SSIM}). However, when assessed for the visual perceptional metric FID, TES-GAN achieved the best FID, followed by other transformer based GANs and the non-GAN model SwinMR.

Training time in the TABLE~\ref{tab:FID} refers to the model training time for 100 steps on two NVIDIA 3090 GPUs with 24GB GPU RAM. Inference time refers to the model inference time for one image on an NVIDIA 3090 GPU with 24GB GPU RAM. For training time, dual-discriminator GAN based transformers, i.e., EES-GAN and TES-GAN, have the longest training time, followed by single-discriminator GAN based transformers ST-GAN.

\section{DISCUSSION}

In this work, we have assessed the performance of the transformer models for fast MRI reconstructions. Standalone Swin transformer model (SwinMR) and its GAN based variants, i.e., ST-GAN (standard GAN), EES-GAN and TES-GAN (dual-discriminator structure with edge and texture enhanced) have been evaluated, against other CNN based GANs (DAGAN and PIDD-GAN). Experiment results have shown that all transformer models have achieved the best performance. SwinMR has outperformed other CNN based methods with higher SSIM and PSNR.

To further explore the capabilities of transformers on the MRI reconstruction, we coupled the Swin transformer with a U-Net based adversarial discriminator, forming ST-GAN. 
In order to further enhance the edge and texture, an additional U-Net discriminator using the edge information by Sobel or the texture information by Gabor, was appended to ST-GAN to form EES-GAN or TES-GAN respectively. 

\begin{figure}[thpb]
  \centering
  \includegraphics[scale=0.8]{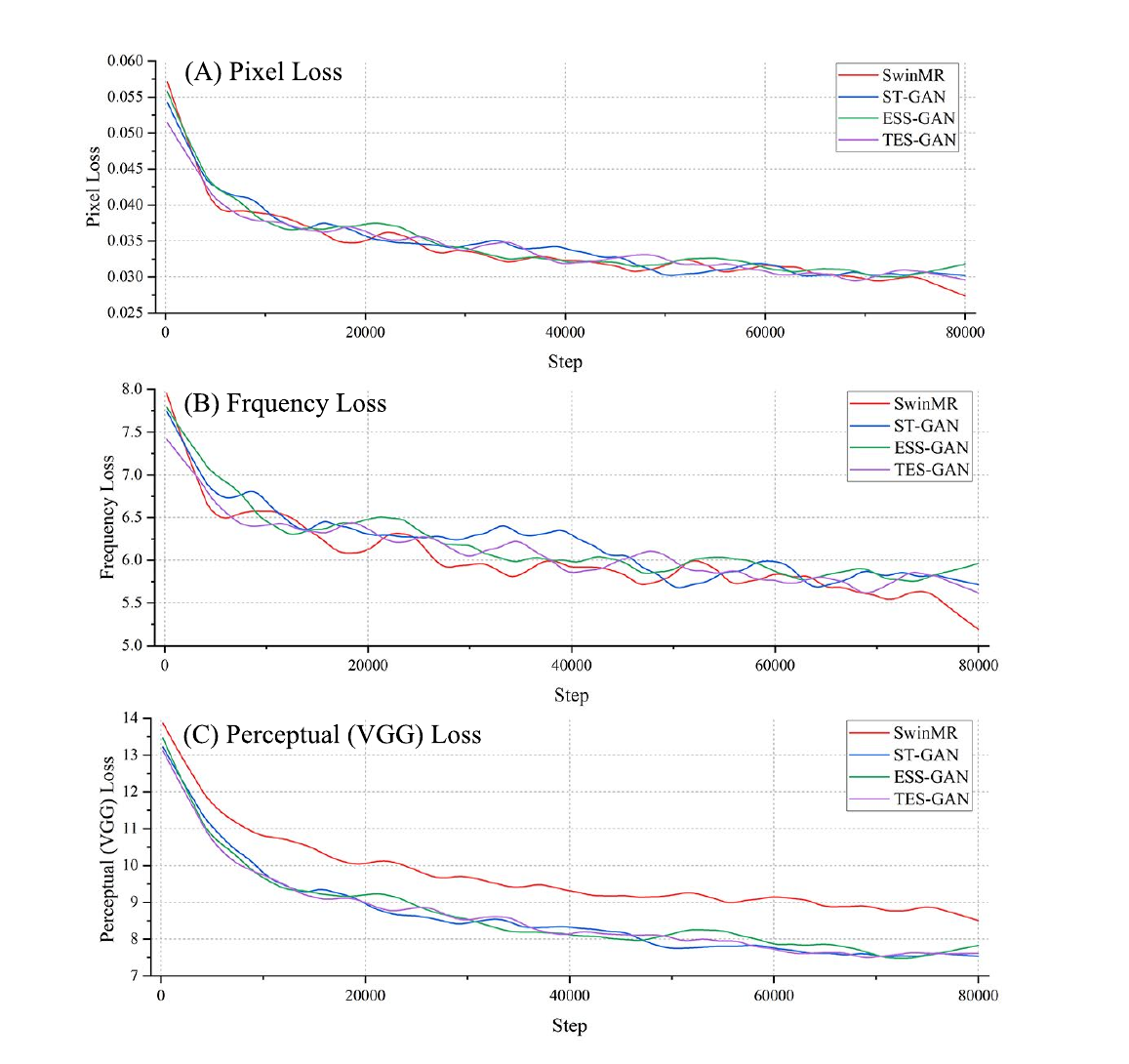}
  \caption{
  Model learning curves for SwinMR, ST-GAN, ESS-GAN and TES-GAN with respective to (A) Pixel-wise L1 loss, (B) Frequency L1 loss, and (C) Perceptual VGG loss.
  }
  \label{fig:loss_function}
\end{figure}

We initially hypothesised that the adversarial training in ST-GAN, EES-GAN and TES-GAN can further improve the quality of reconstruction beyond SwinMR. The reconstruction quality was then assessed by the PSNR and SSIM metrics depicting pixel-to-pixel and structural similarities and the FID score for the visual perceptual experience in a higher dimension. Although richer textures and edges seemed to have been restored by these transformer based GANs (Fig.~\ref{fig:comparison_sample} (B)) with lower FID scores (TABLE~\ref{tab:FID}) and much lower converged perceptual VGG loss in training (Fig.~\ref{fig:loss_function}), these models exhibited lower PSNR and SSIM scores than the standalone SwinMR's (Fig.~\ref{fig:SSIM}). This means that although the model may give better visual perceptional experience in terms of its texture (illustrated by FID and converged perceptual VGG loss in training), the images reconstructed exhibited difference when compared in a pixel-to-pixel strategy (illustrated by PSNR and SSIM).

Such paradoxes may bring some flaws in the reconstructed images. Although all GAN generated images in Fig.~\ref{fig:comparison_sample} (B) seemed clear, when compared to ground truth we can see some hallucinated brain structural textures were added to the images. This may suggest that these GAN based reconstruction methods could have less specificity when reconstructing brain MRI, particularly when the acquisition is greatly undersampled at only 10\% rate. Previous studies have proven that different from $l_1$ or $l_2$ loss which focuses on the reconstruction of global low-frequency structures, adversarial loss in GANs focuses on generating high-frequency details~\cite{Isola2016,yuan2020sara,lv2021pic}. This further explains the clear yet incorrect high-frequency textures generated when the MRI is greatly undersampled (e.g., using G1D10\%).

In terms of the computational cost, the additional edge/texture enhancing adversarial discriminators in EES-GAN and TES-GAN have greatly increased the training times yet with similar inference times (TABLE~\ref{tab:FID}), while bringing little improvement in terms of the quantitative metrics (TABLE~\ref{tab:FID} and Fig.~\ref{fig:SSIM}) and their converged training losses (Fig.~\ref{fig:loss_function}). Therefore, overall we would like to recommend ST-GAN for undersampled MRI reconstruction for G1D30\%, because the model has enhanced the textures while maintaining its specificity. When with lower undersampling rates (e.g., 10\%), these GAN based transformer models may bring misleading structural changes in the images, where SwinMR is a better choice.

\section{CONCLUSIONS}

Our study has explored the potential of transformers and their GAN variants in fast MRI. We have proposed the standard GAN coupled transformer ST-GAN and dual-discriminator GAN based transformers EES-GAN and TES-GAN for edge and texture enhancement. 
We have assessed their performances by the shallow metrics PSNR and SSIM and the visual perceptional metric FID with a higher dimension. Upon comparison, we can recommend ST-GAN for the reconstruction of undersampled MRI with sampling rates higher than 30\% due to its capability of edge and texture enhancement. When at lower undersampling rates (e.g., 10\%), these GAN based transformer models may give misleading texture enhancement with a lower specificity, where SwinMR is a better choice. In summary, transformer based models have shown great performance in fast MRI, and we can envisage a further development for clinical specific problems and combination with prior information of MR physics.

\addtolength{\textheight}{-7cm}   





\bibliographystyle{IEEEtran}
\bibliography{Trans_GAN_intro}

\end{document}